\documentstyle[preprint,aps]{revtex}

\tighten
\begin{document}

\title{
            Boson-fermion mapping and dynamical supersymmetry
                        in fermion models
}
\medskip

\author{
        P. Navr\'atil$^{a,b}$\footnote{On leave of absence
                   from the
                   Institute of Nuclear Physics,
                   Academy of Sciences of the Czech Republic,
                   250 68 \v{R}e\v{z} near Prague,
                     Czech Republic.},
    H.B. Geyer$^{a,c}$,  and J. Dobaczewski$^{a,d}$
  }

\medskip

\address{
                  $^{a}$Institute of Nuclear Theory,
                  University of Washington, Seattle, WA 98195
\newline
                   $^b$Department of Physics,
                   University of Arizona,
                   Tucson, Arizona 85721
                   \newline
       $^c$Institute of Theoretical Physics, University of
Stellenbosch,
       Stellenbosch 7600, South Africa \newline
       $^d$Institute of Theoretical Physics, Warsaw
       University,
       Ho\.{z}a 69, PL-00-681 Warsaw, Poland}

\maketitle

\bigskip

\begin{abstract}
We show that a dynamical supersymmetry can appear in a purely
fermionic system.  This ``supersymmetry without bosons" is
constructed by application of a recently introduced
boson-fermion Dyson mapping from a fermion space
to a space comprised of collective bosons and ideal fermions.
In some algebraic fermion models
of nuclear structure, particular Hamiltonians may lead to
collective
spectra of even and odd
nuclei that can be unified using the dynamical
supersymmetry concept with
Pauli correlations  exactly taken into account.
\end{abstract}

\bigskip
\bigskip
\bigskip

\narrowtext



\section{Introduction}
\label{sec1}

In its original inception supersymmetry pertains to a system of
fermions and bosons exhibiting an invariance with respect to exchange
between these two classes of particles.  It may therefore be somewhat
surprising to discover that a fermion system on its own can also
exhibit {\em dynamical supersymmetry}. States with even and
odd fermion numbers are then unified in a single representation
of a supergroup. This is discussed in general
below and demonstrated for a specific nuclear model.

States of even and odd nuclei are in principle
eigenstates of the same Hamiltonian obtained for different particle
numbers.  Although there is no fundamental difference between even and
odd nuclei from this point of view, their properties are, however,
quite different.  A unification of spectra of even and odd nuclei into
a single framework is therefore a challenging possibility, with the
prospect of unveiling a basic underlying symmetry.

The notion of supersymmetry has in fact proved to be fruitful in
nuclear structure physics\cite{Iac80}.  Properties of some neighboring
even and odd nuclei can namely be classified and understood in terms
of an assumed supersymmetry within the framework of the interacting
boson-fermion model (IBFM)\cite{IvI91}.
Although this phenomenological supersymmetry does not
necessarily imply dynamical supersymmetry on the microscopic level,
 the IBFM\cite{IvI91} does achieve
a unification of even and odd states on the phenomenological
level. Starting from a common boson-fermion Hamiltonian one
finds that in some instances states of an even nucleus, described by
many-boson wave functions, are linked by supersymmetry to states in a
neighboring odd nucleus in which the odd fermion is treated
explicitly.  It is important to realize, however, that Pauli
correlations between the odd particle and even core are not fully
taken into account in the IBFM.  In this sense the link between
observed supersymmetry and an underlying microscopy is tenuous and
 no detailed microscopic understanding of the IBFM
hypothesis in fact exists so far.

In this paper we investigate the link between dynamical
supersymmetry and some fermion algebraic models.  For this purpose we
apply the recently developed Dyson boson-fermion mapping \cite{NGD95}
which has the key property of mapping an even fermion
system (comprised of
collective pairs) to a purely boson system, and at the
same time an odd fermion system (comprised of
collective pairs and an odd fermion) to a system of
collective bosons and a single ideal fermion.
The construction is valid for any
number of collective pairs and non-collective fermions treated
individually, and preserves the Pauli principle exactly.

The paper is organized as follows. In Sections \ref{sec6} and
\ref{sec2}
we discuss respectively the
general properties of  dynamical supersymmetry
 as it is manifested
in nuclear physics, and the boson-fermion
mapping which we use to construct it.
Sec.\ \ref{sec3} presents a simple example of the appearance of a
dynamical supersymmetry in the SU(2) seniority model.  Then, in Sec.\
\ref{sec4}, we describe a more realistic system with the algebraic
structure of SO(8)$\otimes$SO(5) which also exhibits a dynamical
supersymmetry similar to that appearing in the IBFM.  Conclusions are
drawn in Sec.\ \ref{sec5}.
The present paper is an expanded version of a short
unpublished communication \cite{NGD94}.

\section{Supersymmetric unification of states in even and odd nuclei:
phenomenology vs microscopy} \label{sec6}

The IBFM\cite{IvI91} generalizes the phenomenological interacting
boson model (IBM)\cite{IA87} to odd systems by coupling single fermion
degrees of freedom to the IBM bosons, {\it assuming} these fermions
to be kinematically independent from the bosons.  The most general
one- plus two-body IBFM Hamiltonian contains pure boson and pure
fermion parts, as well as an interacting part, the only requirement
being that boson and fermion numbers are conserved separately.

Even for the most restricted fermion subspaces this Hamiltonian still
contains far too many parameters to serve as the basis for a purely
phenomenological analysis.  Microscopically inspired considerations
can, on the one hand, serve as a guideline to effect a reduction to a
few manageable
terms\cite{IvI91}, while dynamical bose-fermi symmetry
and, eventually, dynamical supersymmetry can also serve to restrict
the dynamics.

These dynamical symmetries, their consequences and eventual comparison
with experimental data are discussed in detail in Refs.\
\cite{IvI91,BBI81,Bal82,Ver87}. (See also Ref.\ \cite{JG96} for a recent
analysis where (extended) supersymmetry is successfully used to
account for properties of negative parity states in
$^{194}$Ir.) Here
we only briefly recount the most salient aspects of, respectively,
dynamical bose-fermi and supersymmetry on the phenomenological level
in order to make it clear what we accomplish in our construction of a
dynamical supersymmetry on the microscopic level.

Recall that dynamical symmetry refers to a Hamiltonian which can be
written in terms of Casimir invariants of the algebras appearing in a
{\it single} chain of associated subgroups only, whence an analytical
expression can be found for the energy spectrum by exploiting a basis
labelled according to irreps of the highest group in the
chain\cite{IvI91,IA87}.  This situation implies various relationships
between strength parameters of the general Hamiltonian and thus
represents a considerable restriction.  From the phenomenological
point of view one studies whether dynamical symmetries indeed appear
among nuclear spectra; from the microscopic point of view one would
endeavour to motivate a set of strength parameters conforming to a
dynamical symmetry.  An extensive literature exists for both types of
study -- see e.g.  Ref.\ \cite{IvI91,IA87} for a presentation of
phenomenological studies and Ref.\ \cite{KM91} for a discussion of
microscopic investigations.

The general form of the highest or first group in the subgroup chain
pertaining to a bose-fermi symmetry of the IBFM type can be written as
G$_{\rm B}\otimes$G$_{\rm F}$
and in the simplest version of the IBM with one $s$-
and five $d$-bosons this takes the form U(6)$\otimes$U(2$\Omega$)
with
$2\Omega$ the dimension of the fermion space. Dynamical symmetries
for situations where the boson and fermion subchains are essentially
independent are the least restrictive and not very interesting.
However, when some of the algebras appearing in the two chains are
the same or isomorphic, one has the possibility to couple them to a
spinor algebra. When this happens beyond the level of the rotation
group, the dynamical structure seems to be sufficiently restrictive to
lead to interesting and manageable
classification schemes which are
also realised in some nuclei\cite{IvI91,BBI81,Bal82,Ver87}.

It is important to realize that a dynamical bose-fermi symmetry with a
given set of strength parameters pertains to one specific odd
nucleus, the number of bosons $N_{\rm B}$ being fixed as
$N_{\rm B}=\frac{1}{2}(n-1)$
with $n$ the number of valence nucleons.

A more ambitious attempt to implement dynamical symmetry is to
consider U($m_{\rm B}$)$\otimes$ U($2\Omega$), with $m_{\rm B}$
the number of single boson states, as being embedded in the
supergroup U($m_{\rm B}/2\Omega$).
This implies that one is considering the
classification of nuclear states in a set of neighbouring even and odd
nuclei within a single supersymmetric representation $[\aleph\}$,
where
$\aleph= N_{\rm B}+{\cal N}$, with ${\cal N}$ the number of unpaired
nucleons, and where the Hamiltonian parameters and all other
parameters appearing in physical operators are the same for all the
nuclei under consideration -- clearly a severe dynamical restriction.

Note that the supersymmetric representation is specified not by the
effective number of fermions, $2N_{\rm B}+{\cal N}$,
but by the number of bosons
plus fermions, $N_{\rm B}+{\cal N}$,
which of course underlies the possibility to
use the single supersymmetric representation for a set of adjacent
nuclei.  Although this type of dynamical supersymmetry is invariably
found to be broken, there seems to be enough evidence
\cite{IvI91,BBI81,Bal82,Ver87,JG96} that it is a useful scheme for
classification and calculation of nuclear properties for a number of
nuclei.

What concerns us mostly in the remainder of this paper is the question
whether dynamical supersymmetry, as summarized above, can be
compatible with the Pauli principle or, alternatively, if dynamical
supersymmetry can be an exact property of a fermion system.  From the
point of view that there are important Pauli corrections to the lowest
order association between collective fermion pairs and IBM
bosons\cite{KM91}, one might anticipate a negative answer to this
question.  Nevertheless, we show that the implementation of
appropriate boson-fermion mappings indeed reveals instances where this
compatibility holds.  These mappings, introduced and refined in Refs.\
\cite{DSG93,NGD95}, and discussed and exploited below, introduce an
equivalence between a system of interacting fermions on the one hand,
and on the other a system of interacting bosons and fermions which are
{\it by construction kinematically independent}.  Although no
guarantee in itself, this clearly fulfills the minimum requirement for
a dynamical supersymmetry to exist, namely to have the appropriate
degrees of freedom.  Moreover, as elaborated below, one can indeed
find instances where dynamical supersymmetry emerges as an exact
classification scheme of fermion states.  This is simply an
alternative, but equivalent, classification to whatever classification
scheme may have been adopted on the fermion level.  What the
boson-fermion mapping accomplishes is to make the inherent
supersymmetric nature transparent
in these instances, although they
should also be identifiable on a purely algebraic level in terms of
relationships between standard and supersymmetric representations.

Apart from providing a concrete link between fermion dynamics and
dynamical supersymmetry, the use of boson-fermion mappings also allows
one to {\it construct} various transition operators appropriate to the
boson-fermion description.  This is in contrast to the
phenomenological situation where one is obliged to truncate an
infinite series of combinations of boson and fermion operators with
phenomenological parameters and terms only restricted by their tensor
and particle number changing properties\cite{IvI91,IA87}.  It should
be emphasized that the choice of these transition operators in
phenomenological models such as the IBM or IBFM is {\it not} dictated
by the Hamiltonian parameters in general, specifically also in the
case of dynamical symmetry or supersymmetry.  (See also Ref.\
\cite{GND94} for a discussion of this point.)  In the present context
we show e.g.\ how the single fermion transfer operator is uniquely
specified in the boson-fermion description by the appropriate mapping
and how it compares with phenomenological results.

Finally, our work provides the possibility of a direct link
between phenomenological models which introduce dynamical
supersymmetry in an inherently fermion problem, such as is in the
IBFM, and fermion models which exploit a similar structure by
considering both vector and spinor representations of the the overall
symmetry group, such as in the very recent fermion
dynamical symmetry model (FDSM) analysis\cite{PPF96}.
Our formalism is thus tailor made to investigate the similarities and
differences between the two approaches.

\section{Dyson boson-fermion mapping of the collective algebra}
\label{sec2}

In order to keep the paper as self-contained as possible, we briefly
retrace the basic steps in the derivation of the Dyson boson-fermion
mapping \cite{NGD95} used in subsequent sections.
We start from a fermion
collective pair algebra comprised  of collective fermion pair creation
operators $A^j$=$\textstyle{\frac{1}{2}} \chi^j_{\mu\nu} a^{\mu}
a^{\nu}$, pair annihilation operators $A_i$, and their commutators
$[A_i,A^j]$.  The notation exploits a summation convention in which
upper (lower) indices denote creation (annihilation) operators, as in
$a^{\mu}$=$a^+_{\mu}$ and $A^j$=$A^+_j$.  We also
assume\cite{DSG93,DGH91} that the collective fermion operators obey
the algebraic closure relations
$   \left[\left[A_i,A^j\right],A_k\right] = c^{jl}_{ik}A_l ,$
i.e., operators $A^j$, $A_i$, and $[A_i,A^j]$ are assumed to form
a collective spectrum generating algebra\cite{SGA88}.
These relations guarantee an exact decoupling of the even collective
space $ |\Psi_{\rm{even}}\rangle=A^iA^j\ldots{}A^k|0\rangle$ from all
other even fermion states.  Similarly, by adding an odd fermion, one
obtains an exactly decoupled odd collective space
$|\Psi_{\rm{odd}}\rangle=a^{\mu}A^iA^j\ldots{}A^k|0\rangle$.

In a boson-fermion description an even state
$|\Psi_{\rm{even}}\rangle$ should be represented by an ideal space
state $|\Psi_{\rm{even}}$), say, which contains {\it bosons only},
with the odd ideal space states $|\Psi_{\rm{odd}})$ containing an
additional fermion $\alpha^{\mu}$.  Following traditional terminology
this is also referred to as an ideal fermion\cite{KM91}.  By
definition it commutes with all boson operators,
$[\alpha^{\mu},B_i]$=$[\alpha^{\mu},B^i]$=0.  The required
representation in the ideal space can be achieved by constructing an
appropriate boson-fermion mapping (see the Appendix),
\begin{mathletters}
\label{e228}\begin{eqnarray}
   A^j               &\longleftrightarrow& R^j - {\cal{A}}^j =
                       B^i\big[{\cal{A}}_i,{\cal{A}}^j\big]
                     - \textstyle{\frac{1}{2}}c^{jl}_{ik}B^i B^k
B_l
                                          , \label{e228c} \\
   \big[A_i,\!A^j\big] &\longleftrightarrow&
                            \big[{\cal{A}}_i,{\cal{A}}^j\big]
                    - c^{jl}_{ik}B^kB_l
                                              , \label{e228b} \\
   A_j               &\longleftrightarrow&
                            B_j               , \label{e228a} \\
   a^{\nu}           &\longleftrightarrow&
                            X^{-1}\left(\alpha^{\nu}
                    +
B^i\big[{\cal{A}}_i,\alpha^{\nu}\big]\right)X
                                              ,  \label{e228e} \\
   a_{\nu}           &\longleftrightarrow&
                            X^{-1}\alpha_{\nu}X      ,
\label{e228d}
   \end{eqnarray}
\end{mathletters}%
where ${\cal{A}}^j$= $\textstyle{\frac{1}{2}}\chi^j_{\mu\nu}
\alpha^{\mu}\alpha^{\nu}$ are collective pairs of {\em ideal
fer\-mions}.
The similarity transformation $X$ has the explicit form
\begin{equation}\label{eq3}
     X = \sum_{n=0}^{\infty} \bigl(\frac{1}{C_{\rm F}-\widehat{C}_{\rm F}}
        {\cal A}^i B_i\bigr)^n
        {\textstyle \raisebox{-2ex}{$\widehat{}$}} \quad ,
\end{equation}
where $C_{\rm F}$=${\cal{A}}^k{\cal{A}}_k$ is an operator which leaves
invariant the ideal-fermion core subalgebra composed of operators
$[{\cal{A}}_i,{\cal{A}}^j]$, i.e.,
$\left[C_{\rm F},[{\cal{A}}_i,{\cal{A}}^j]\right]$=0.  Here we use the
notation\cite{Gey86,KV87} of a deferred-action operator
$\widehat{C}_{\rm F}$ which should be evaluated at the position indicated by
``$\:$\raisebox{-1ex}{$\widehat{}$}$\:$''.  Equivalently, one can
write Eq.\ (\ref{eq3}) by using multiple sums over eigenstates of
$C_{\rm F}$, in which case $1/(C_{\rm F}-\widehat{C}_{\rm F})$ can be
replaced by typical energy denominators.

We can now discuss the structure of the images of single fermion
operators, Eqs.\ (\ref{e228e}) and (\ref{e228d}).  An explicit general
evaluation of these images is difficult because the operators
$\alpha^{\nu}$ and $\alpha_{\nu}$ do not commute with the
ideal-fermion invariant operator $C_{\rm F}$ and one has to consider
branching rules of the collective algebra.  Before presenting
solutions in particular cases we can, however, analyze some general
properties of these images.

We note that the similarity transformation (\ref{eq3}) does not
change
any state in which there are no bosons $B^j$, and therefore the
single-fermion states are mapped onto single ideal fermion
states,
$a^{\nu}|0\rangle\leftrightarrow\alpha^{\nu}|0)$.  Since the
images of
pair creation operators (\ref{e228c}) do not change the ideal
fermion
number, collective odd states $ |\Psi_{\rm{odd}}\rangle$ are
mapped onto ideal states with one ideal fermion only.

The two-fermion states are mapped as
$a^{\mu}a^{\nu}|0\rangle\longleftrightarrow
\bigl(\alpha^{\mu}\alpha^{\nu}+\chi^{\mu\nu}_iB^i
-\frac{1}{C_{\rm F}}\chi^{\mu\nu}_i{\cal{A}}^i\bigr)|0)$ and in general
contain the non-collective pair of ideal fermions
$\alpha^{\mu}\alpha^{\nu}|0)$.  However, when the collective pair
$A^j$ is formed by summing the pairs $a^{\mu}a^{\nu}$ with collective
amplitudes $\textstyle{\frac{1}{2}} \chi^j_{\mu\nu}$, the ideal
non-collective pairs above recombine (note that
$C_{\rm F}{\cal{A}}^j|0)$=$g{\cal{A}}^j|0)$) and only the boson state
$gB^j|0)$ remains.  Again, since the images (\ref{e228c}) conserve the
ideal fermion number, the same recombination mechanism is also valid
for any even state.

Similarly as in the even case, spurious states may
appear in spectra of mapped operators when diagonalized
in the complete ideal boson-fermion space.  However, they do not
contaminate physical states and can be identified after the
diagonalization.
The methods which can be used to this effect are
analogous to those discussed in Ref.~\cite{DGH91}, and will not be repeated
here.

\section{The SU(2) seniority model and dynamical supersymmetry}
\label{sec3}

The textbook SU(2) seniority model \cite{RS80} is very often used in group
theory models to illustrate  the key aspects of the methods
used. Although nothing new can be said about the model itself,
its simplicity and intuitiveness allows for a very suitable
test ground where advanced approaches can be explained in simple
terms. In this Section we apply to this model our boson-fermion
mapping and discuss the resulting dynamical supersymmetry.

The SU(2) model is defined by considering in  a single-$j$
shell the monopole pair creation operator
\begin{equation}\label{spair}
S^+ = \sqrt{\frac{\Omega}{2}} (a^+_j a^+_j)^{(0)}
\; ,
\end{equation}
 with $\Omega=j+\textstyle{\frac{1}{2}}$.
 It fulfills the commutation relation
\begin{equation}\label{coms}
\left[S,S^+\right] = \Omega - n ,
\end{equation}
where $n$ is the fermion number operator
in the single-$j$ shell. The SU(2) algebra
can
be generalized to describe odd systems by constructing
the superalgebra  generated by the operators $S^+$,
$S$, $\Omega$$-$$n$,
$a_{jm}$, and $a_{jm}^+$. The
relevant
commutation relation is
\begin{equation}\label{comas}
\left[a_{jm}^+,S\right] = -\tilde{a}_{jm} \; ,
\end{equation}
with $\tilde{a}_{jm}=(-1)^{j-m} a_{j,-m}$,
while the single-fermion operators obey the standard anticommutation
relations.  Clearly this is a rather trivial superalgebra as the
elements of the odd sector (single fermion operators)
anti-commute only to the identity. Alternatively, by considering
the {\it commutator} of single-fermion operators, the set of bi- and
single-fermion operators may of course also be viewed as generators of
a standard (orthogonal) algebra.

In the single-$j$ shell we consider the
pairing Hamiltonian
\begin{equation}\label{pairh}
H=-G S^+ S \; ,
\end{equation}
which has the energy spectrum \cite{RS80}
\begin{equation}\label{senen}
E=-\textstyle{\frac{1}{4}}G (n-v)(2\Omega-n-v+2) \; ,
\end{equation}
with the seniority quantum number $v$ denoting the number of
fermions
not coupled to angular momentum zero.
This Hamiltonian describes both the even and odd systems,
and the spectra in both cases are given by the same expression
(\ref{senen}) with $v$ even or odd, respectively.

We can apply to this model
the Dyson boson-fermion mapping described in Sec.\ \ref{sec2}
to find an equivalent description in the
boson-fermion space.  For the SU(2) algebra,
the operator $C_{\rm F}$ depends on the
ideal fermion number operator ${\cal N} =
\sum_m\alpha^\dagger_{jm}\alpha_{jm}$ only,
and one may derive an explicit
form of the similarity transformation\cite{NGD95}.  As we have
\begin{equation}\label{eqsu1}
C_{\rm F} - \widehat{C}_{\rm F} =
\case{1}{2}({\cal N}-\hat{\cal N})(\Omega+1-\case{1}{2}({\cal
N}+\hat{\cal N})) ,
\end{equation}
the transformation (\ref{eq3}) is
\begin{equation}\label{eqsu2}
 X = \frac{(\Omega-\case{1}{2}({\cal N}+\hat{\cal
N}))!}{(\Omega-\hat{\cal N})!}
     {\rm exp}\left[{\cal S}^{\dagger} B \right] \:
{\textstyle \raisebox{-1ex}{$\widehat{}$}}\qquad.
\end{equation}
 Specializing from the general case to SU(2), we have
introduced here the ideal fermion operators
$\alpha^\dagger_{jm}$ and
$\alpha_{jm}$, which commute
with the ideal boson operators $B^\dagger$ and
$B$, and the ideal fermion pair operators
${\cal S}^{\dagger}$ and ${\cal S}$ obtained from
$S^+$ and $S$ by
replacing $a$ by $\alpha$.  The general Dyson
boson-fermion mapping (\ref{e228})
is  then obtained for the SU(2) case
in the
form
\widetext
\begin{mathletters}\label{sulmap}
\begin{eqnarray}
S^+ &\longleftrightarrow&
                \Omega B^\dagger - B^\dagger B^\dagger B
-B^\dagger {\cal N}
   = B^\dagger \left(\Omega - N_{\rm B} - {\cal N}\right)
   = B^\dagger \left(\Omega - \aleph \right)
                                              , \label{su141c} \\
S  &\longleftrightarrow&
                       B                    , \label{su141a} \\
n &\longleftrightarrow&
                        2 B^\dagger B +{\cal N}
                      = 2 N_{\rm B} + {\cal N}
                      = \aleph + N_{\rm B}
                                              , \label{su141b} \\
a^+_{jm} &\longleftrightarrow&
                       \alpha^{\dagger}_{jm}\frac{\Omega-\aleph}
                       {\Omega-{\cal N}}
                 + B^\dagger{\tilde\alpha}_{jm}
                 - {\cal S}^{\dagger} {\tilde\alpha}_{jm}
     \frac{\Omega - \aleph}{(\Omega-{\cal N})(\Omega-{\cal N}+1)}
                                             , \label{su141e} \\
a_{jm} &\longleftrightarrow&
                             \alpha_{jm}
 + {\tilde\alpha}^{\dagger}_{jm} B \frac{1}{\Omega-{\cal N}}
                     + {\cal S}^{\dagger} \alpha_{jm} B
 \frac{1}{(\Omega-{\cal N})(\Omega-{\cal N}+1)}
                                              , \label{su141d}
\end{eqnarray}\end{mathletters}\narrowtext%
where $\aleph= N_{\rm B}+{\cal N}$.
We see that the single fermion images (\ref{su141e}) and
(\ref{su141d}) are finite and contain terms changing the ideal fermion
number by one only.  Furthermore, they preserve
{\em exactly} the anti-commutation
relations on the full ideal space, i.e., as operator identities.
This is guaranteed by our construction method, and
can be verified by explicit calculation.
The preservation of the commutation and anticommutation relations
ensures the exact preservation of the Pauli exclusion principle,
once the original fermion problem is mapped into the boson-fermion
space.

Clearly, the mapping (\ref{sulmap})
transforms the 2-body Hamiltonian
(\ref{pairh})  into a 1-plus-2-body boson-fermion
Hamiltonian of the form
\begin{equation}\label{pairhbf}
H_{\rm BF} = -G N_{\rm B} (\Omega - N_{\rm B} +1 - {\cal N}) \; .
\end{equation}
Hamiltonians (\ref{pairh}) and (\ref{pairhbf}) have exactly the
same spectrum (\ref{senen}).
Hamiltonian (\ref{pairhbf}) can also be expressed in a form
which stresses its dependence on the total number of bosons and fermions
$\aleph$, i.e.,
\begin{equation}\label{pairhbfa}
H_{\rm BF} = -G (\aleph - {\cal N}) (\Omega +1 - \aleph) \; .
\end{equation}
Note also that the boson-fermion interaction term in (\ref{pairhbf}),
which reads $GN_{\rm B}{\cal N}$,
can be expressed in terms of the odd generators,
$O^{\dagger}_m=\alpha^{\dagger}_{jm}B$ and $O_m=B^{\dagger}\alpha_{jm}$,
of the U(1/2$\Omega$) superalgebra.
Since the boson and ideal fermion number operators can be
linked to even generators (see also expression (\ref{bosfch}) and its
discussion below),
this identification makes it possible to write the
Hamiltonian (\ref{pairhbf})
in yet another form in terms of both
even generators and {\it supergenerators} of U(1/2$\Omega$):
\begin{equation}\label{pairhbfb}
H_{\rm BF} = -G \left( N_{\rm B} (\Omega - N_{\rm B} +1)
                       + {\cal N} - \sum_m O^{\dagger}_m O_m \right) \; .
\end{equation}

After the mapping states with a given fermion number $n$ are
classified according to the number of bosons  $N_{\rm B}$
and the number of ideal fermions  $\cal N$, these numbers still
being related to the fermion number $n$ through
$n$=$2N_{\rm B}$+${\cal N}$. The number of ideal fermions
 now corresponds to the seniority quantum number, because
all fermion pairs coupled to zero angular momentum are mapped
onto bosons.  This latter
identification follows by construction, as any attempt to create an
ideal fermion pair of angular momentum zero by applying (\ref{su141e})
successively (with appropriate coupling), results in  the
creation of the boson  $B$.
As long as we do not overfill the $j$ level by
too many particles, i.e.,
restrict the ideal space to physical states
with $2N_{\rm B}+{\cal N}\leq 2\Omega$, the fermion space results are
exactly reproduced in the boson-fermion space.

In general it is not possible, as in this simple case, to make, for a
given fixed particle number, an {\it a priori} selection of simple
ideal space states which will span the physical subspace.  However, as
elaborated in Ref.\ \cite{DGH91}, one can
exploit the simplicity
of the ideal space basis for the diagonalization
of the mapped Hamiltonian. The physical
eigenstates do not mix with the unphysical states and can
by identified after the diagonalization.
(See Ref.\ \cite{NGDD95} for a recent application of this procedure.)

In the original fermion space of the SU(2) seniority
model, the
Hamiltonian eigenstates are classified according
to the representations of the subgroup chain
\begin{equation}\label{fermch}
{\rm SO(}4\Omega{\rm+1)}\supset {\rm SU(2)}\otimes {\rm
Sp(}2\Omega{\rm )}
\supset {\rm U(1)}\otimes {\rm SO(3)} \; ,
\end{equation}
while in the boson-fermion space they are classified by
the quantum numbers
derived from the chain
\begin{equation}\label{bosfch}
{\rm U(}1/2\Omega{\rm )}\supset
{\rm U_{\rm B}(1)}\otimes {\rm U_{\rm F}(}2\Omega{\rm )}\supset
{\rm U_{\rm B}(1)}\otimes {\rm U_{\rm F}(}1{\rm )}\otimes
{\rm Sp_{\rm F}(}2\Omega{\rm )}\supset
{\rm U_{\rm B}(1)}\otimes {\rm U_{\rm F}(}1{\rm )}\otimes
{\rm SO_{\rm F}(3)} \; .
\end{equation}
In the second chain  the supergroup U(1/2$\Omega$) is generated by
$B^\dagger B$, $\alpha^\dagger_{jm}\alpha_{jm'}$,
$\alpha_{jm}^{\dagger} B$, and $B^\dagger\alpha_{jm}$,
with the first two operators belonging to the even sector
and the remaining two to the odd sector of the superalgebra, respectively.
The appearance of U(1/$2\Omega$) in the chain is
clearly suggested by the form (\ref{pairhbfb}) and also dictates
that the same Hamiltonian
parameters in (\ref{pairhbf}) are used for both even and odd states,
if the Hamiltonian should be viewed as a phenomenological
boson-fermion Hamiltonian. From the mapping point of view there is of
course no real choice in this matter, as the original fermion
Hamiltonian pertains to any number of particles.
Nevertheless, it is interesting to note already here how the
phenomenological extension to superalgebras {\it \`a la} IBFM may be
suggested from a microscopic point of view.

In the dynamical supersymmetry concept,  those states of the
system with the
same number of  {\it ideal} particles
$\aleph$ belong to a single supersymmetric representation
$[\aleph\}$
of the supergroup, here U(1/$2\Omega$).
(Note that $\aleph$=$N_{\rm B}+{\cal N}$ 
differs from the fermion
number $n$ and its ideal space image (\ref{su141b}),
$n \longleftrightarrow 2 N_{\rm B} +{\cal N})$.
Starting from, say, the state with no fermions $|N_{\rm B})$, the
other
states of the multiplet are obtained by successive
application of
supergenerators $\alpha_{jm}^{\dagger} B$.

Although  one can embed all states of the SU(2) model in a
sequence of representations of the supergroup U(1/$2\Omega$),
this does not really provide other interesting
physical consequences. This is so because in the chain (\ref{bosfch})
the supergroup U(1/$2\Omega$) is immediately split into the
boson and fermion sectors which remain separate  down to the
bottom of the chain.  As discussed in Sec.\ \ref{sec6},
potentially interesting situations which are restrictive from the
supersymmetry point of view, occur if at a
certain level  of subgroups one may find the same
subgroups in the boson and
fermion sectors, and combine them together into a given boson-fermion
subgroup. Such a situation  arises in a fermion
model with a
richer structure than the present SU(2) case and is discussed in
the next section.
 In terms of algebras appearing in the chain (\ref{bosfch}),
we here only find a trivial example
of the boson subalgebra U$_{\rm B}$(1),  generated
by the boson number operator $N_{\rm B}$, and the fermion
subalgebra U$_{\rm F}$(1), generated by the fermion number
operator ${\cal N}$, which can be combined into the boson-fermion
subalgebra U$_{\rm BF}$(1),
generated by $\aleph$=$N_{\rm B}$+${\cal N}$.

 At the same time, even the simplest example of the SU(2)
seniority model discussed here provides interesting consequences for the
structure of the  single-fermion transfer operators and the
spectroscopic factors. After the mapping, the single-fermion
operators acquire terms which are responsible for the Pauli correlations
between the even core and the odd particle.
In the phenomenological supersymmetric models these terms are
postulated together with some arbitrary numerical constants, whereas
in the supersymmetric picture derived from the boson-fermion
mapping they are fixed by the mapping procedure itself. For example,
the image of the fermion annihilation operator
(\ref{su141d}) is a combination of the ideal fermion annihilation
operator with two corrective terms. The first
one corresponds to an annihilation
of the same ideal fermion accompanied by an attempt to replace
a boson by a collective ideal fermion pair. This term ensures that
after the annihilation the remaining fermions still obey the
correct statistics. The second  term corresponds to the
standard annihilation of the boson replaced by an appropriate fermion
which has been the partner of the annihilated fermion in a collective pair.
Of course, the relative weights of these corrective terms depend
on how many ideal fermions have already been present before the
annihilation.

To  evaluate and understand different aspects of the
original fermion and the mapped boson-fermion description,
it is instructive to calculate the spectroscopic factors
for the simplest states in the SU(2) model.
In the original fermion space this can be done by
using the commutation relations (\ref{coms}) and
(\ref{comas}), and we
arrive at the
result
\widetext
\begin{equation}\label{spect}
\langle n+1,v=1,j||a^\dagger_j||n,v=0\rangle=
-\langle n,v=0||\tilde{a}_j||n+1,v=1,j\rangle=-\sqrt{2\Omega-n}
\; .
\end{equation}
\narrowtext
To get this relation one has to calculate the normalization factors
of the fermion states.

Evaluating  matrix elements (\ref{spect}) in the mapped
boson-fermion
space is much simpler. In particular, one can work with the ideal
boson-fermion
eigenstates of the Hamiltonian (\ref{pairhbf})
corresponding to the states with ${\cal N}$=0 and 1,
e.g., $|B^{N_{\rm B}})$, and $|B^{N_{\rm B}},\alpha^\dagger_j)$.
Using Eqs.\ (\ref{su141e}) and (\ref{su141d})
we directly read off the results
\begin{mathletters}\label{bfspec}
\begin{eqnarray}
(B^{N_{\rm B}},\alpha^\dagger_j||(a^+_j)_{\rm BF}||B^{N_{\rm B}})&=&
-\sqrt{\frac{2}{\Omega}}(\Omega-N_{\rm B}) \; , \\
-(B^{N_{\rm B}}||(\tilde{a}_j)_{\rm BF}||B^{N_{\rm
B}},\alpha^\dagger_j)&=&
-\sqrt{2\Omega} \; ,
\end{eqnarray}
\end{mathletters}%
where $(a^+_{jm})_{\rm BF}$ and $(\tilde a_{jm})_{\rm BF}$
are the boson-fermion
images of $a^+_{jm}$
and $\tilde a_{jm}$, respectively.
In this calculation, only  simple boson normalization
factors  enter.
After the hermitization  (see Ref.\ \cite{SGH92} and references
therein) and identification of  2$N_{\rm B}=n$, matrix
elements
(\ref{bfspec}) reproduce the correct value of $-\sqrt{2\Omega-n}$.

 It is probably worthwhile to conclude this section by carefully
distinguishing two superalgebraic structures we identified in two
different contexts.  Firstly, the algebraic structure obtained by
extending a collective bifermion algebra with single fermion operators
leads to a trivial superalgebra structure.  Nevertheless, this
identification plays an important role in the construction of
boson-fermion images developed in Refs.\ \cite{NGD95,DSG93}.
Secondly, after mapping of the fermion Hamiltonian, it becomes
possible to analyse the resulting equivalent boson-fermion Hamiltonian
in terms of a subgroup chain with a supergroup as first member, and
classify states of neighbouring even and odd systems according to a
single representation of that supergroup.  This analysis relies on the
identification of the mapped Hamiltonian as being constructed from the
generators of subalgebras of a {\it non-trivial}
superalgebra, generally different
from the one appearing on the original fermion level.

\section{SO(8) and SO(8)$\otimes$SO(5) mapping, and dynamical
supersymmetry} \label{sec4}

In this section we discuss the appearance of dynamical
supersymmetry
in a more complicated model, namely, in  the Ginocchio
SO(8) model \cite{Gin80}.
This model is defined by collective pairs
\widetext
\begin{mathletters}\label{so8gen}
\begin{eqnarray}
   F^+_{JM} &=&
   \sqrt{\textstyle{\frac{1}{2}}} \sum_{j_1j_2}(-1)^{J+i+k+j_1}
   \frac{\hat{j_1}\hat{j_2}}{\hat{k}}
   \left\{\begin{array}{ccc} j_1 & j_2 & J \\
                             i   & i     & k
\end{array}\right\}
   \left(a^+_{j_1}a^+_{j_2}\right)^{(J)}_{M} , \label{so81} \\
   P_{JM} &=&
   -\sqrt{2\Omega} \sum_{j_1j_2}(-1)^{J+i+k+j_1}
   \frac{\hat{j_1}\hat{j_2}}{\hat{k}}
   \left\{\begin{array}{ccc} j_1 & j_2 & J \\
                             i   & i     & k
\end{array}\right\}
   \left(a^+_{j_1}\tilde{a}_{j_2}\right)^{(J)}_{M} , \label{so82}
\end{eqnarray}
\end{mathletters}%
\narrowtext
with $i=\frac{3}{2}$ and $k$ integer. In (\ref{so81}) only
$S^+$ ($J$=0) and $D^+$ ($J$=2)
pairs are allowed, while in (\ref{so82}) $J$ takes values
0, 1, 2, and 3.

In order to generalize the SO(8) algebra to a superalgebra,
we include the
$2\Omega$=(2$i$+1)(2$k$+1) creation and annihilation operators
$a^\dagger_{jm}$ and $a_{jm}$, where $j$=$|k$$-$$i|,\ldots,k$+$i$.
A boson-fermion mapping of this algebra was  derived in
Ref.\cite{NGD95}.
In this case the series defining the similarity
transformation
\begin{equation}\label{eqso86}
 X = \sum_{k=0}^{\infty} (\frac{1}{C_{\rm F} - \widehat{C}_{\rm
F}}
{\cal F}^\dagger_J \cdot \tilde{B}_J)^{k}\:
{\textstyle \raisebox{-1ex}{$\widehat{}$}}\qquad,
\end{equation}
cannot be explicitly summed up because the
denominator
\begin{equation}\label{eqso85}
C_{\rm F} - \widehat{C}_{\rm F} = \frac{1}{\Omega}\left[
\textstyle{\frac{1}{2}}({\cal N}-\hat{\cal N})
(\Omega+6-\textstyle{\frac{1}{2}}({\cal N}+\hat{\cal N})) +
\widehat{C}_{2 {\rm Spin_{\rm F}(6)}}-C_{2 {\rm Spin_{\rm
F}(6)}}\right],
\end{equation}
contains the quadratic Casimir operator of the Spin(6) group
which is not expressible in terms of number operators.
Here ${\cal N}$ is the ideal fermion number operator
${\cal N}=\sum_{jm}\alpha^{\dagger}_{jm}
\alpha_{jm}$
and $C_{2 {\rm Spin_{\rm F}(6)}}=
{\textstyle{1\over 4}}(P_1\cdot P_1+P_2\cdot P_2+P_3\cdot P_3)$.

The similarity transformation $X$ can nevertheless be applied
with
the result that the transformed mapping then reads
\widetext
\begin{mathletters}
   \label{so8images}
\begin{eqnarray}
   F^+_{JM} &\longleftrightarrow& B^{\dagger}_{JM}
      -\frac{2}{\hat{k}^2} \sum_{J_1J_2J_3J'}
       \hat{J_1}\hat{J_2}\hat{J_3}\hat{J'}
          \left\{ \begin{array}{ccc} {i}   & {i}   & J   \\
                                     {i}   & {i}   & J_3 \\
                                     J_2 & J_1 & J'
\end{array}\right\}
             ((B^{\dagger}_{J_1} B^{\dagger}_{J_2})^{(J')}
                   \tilde{B}_{J_3})^{(J)}_{M}  \nonumber \\
   & & + \frac{2}{\hat{i}\hat{k}^2}(-1)^{J_2}\hat{J}_1\hat{J}_2
   \left\{ \begin{array}{ccc} J_1 & J_2 & J \\
                        {i}  &  {i}  & {i} \end{array}\right\}
   \left(B^\dagger_{J_1} {\cal P}_{J_2}\right)^{(J)}_M , \\
   F_{JM} &\longleftrightarrow& B_{JM} , \\
   P_{JM} &\longleftrightarrow& \frac{2\sqrt{2\Omega}}
            {\hat{{k}}}(-1)^{J+2{i}}
            \sum_{J_1J_2}\hat{J}_1 \hat{J}_2
            \left\{ \begin{array}{ccc} J_1 & J_2 & J \\
                               {i}  &  {i}  & {i}
\end{array}\right\}
   (B^{\dagger}_{J_1} \tilde{B}_{J_2})^{(J)}_{M} + {\cal P}_{JM}
,\label{eqso87}
\end{eqnarray}
\end{mathletters}%
for the collective pair operators while for the single-fermion
operators
we obtain
\begin{mathletters}
\label{so8sinim}
\begin{eqnarray}
a^+_{jm} &\longleftrightarrow& \alpha^{\dagger}_{jm}
+ B^\dagger_J \cdot \left[\tilde{{\cal
F}}_J,\alpha^{\dagger}_{jm}\right]
           - \frac{1}{C_{\rm F} - \widehat{C}_{\rm F}}
{\cal F}^\dagger_{J_1}\cdot \tilde{B}_{J_1} \:
\:{\textstyle \raisebox{-1ex}{$\widehat{}$}}\:\:
B^\dagger_J \cdot \left[\tilde{{\cal
F}}_J,\alpha^{\dagger}_{jm}\right]
\nonumber  \\
& & + B^\dagger_J \cdot \left[\tilde{{\cal
F}}_J,\alpha^{\dagger}_{jm}\right]
\frac{1}{C_{\rm F} - \widehat{C}_{\rm F}}
{\cal F}^\dagger_{J_1}\cdot \tilde{B}_{J_1} \:
\:{\textstyle \raisebox{-1ex}{$\widehat{}$}}\:\:\ldots\:\:\:
                                              , \label{so812} \\
  a_{jm} &\longleftrightarrow& \alpha_{jm}
- \frac{1}{C_{\rm F} - \widehat{C}_{\rm F}}
{\cal F}^\dagger_{J_1}\cdot \tilde{B}_{J_1} \:
{\textstyle \raisebox{-1ex}{$\widehat{}$}}\: \: \alpha_{jm}
 + \alpha_{jm} \frac{1}{C_{\rm F} - \widehat{C}_{\rm F}}
{\cal F}^\dagger_{J_1}\cdot \tilde{B}_{J_1} \:
{\textstyle \raisebox{-1ex}{$\widehat{}$}}\:\:\ldots\:\:\:
                                              . \label{so826}
\end{eqnarray}
\end{mathletters}%
\narrowtext

The ideal fermion pair operators
${\cal F}^\dagger_{JM}$ and ${\cal P}_{JM}$ are given by
Eqs.{\ }(\ref{so81}) and
(\ref{so82}), respectively, with the fermion operators $a_{jm}$
replaced by ideal fermion operators $\alpha_{jm}$.
The dots $\dots$ refer to  a class of higher order terms which
will
not contribute to matrix elements between physical states. (See
Ref.
\cite{NGD95} for further discussion.)

A condensate
of collective fermion pairs have now been mapped onto a
condensate of
$s^{\dag}\equiv B^\dagger_0$ and $d^{\dag}\equiv B^\dagger_2$
bosons,
and a condensate of collective fermion
pairs with  one additional
odd fermion, onto a condensate of
$s^{\dag}$ and $d^{\dag}$ bosons
with  one additional
ideal fermion.  This situation is quite reminiscent of the
phenomenological IBFM.
The six bosons above span the symmetric representation of U(6).
The size of the fermion sector depends on
$k$. Let us first discuss the situation with $k$=0 ($j$=3/2). From
Eq. (\ref{eqso87}) we observe that the boson-fermion images
of the multipole operators are
\begin{mathletters}\label{u64}
\begin{eqnarray}
P_{1M} &\longleftrightarrow& 2\sqrt{2}\left[(d^\dagger
\tilde{d})^{(1)}_M
     - \frac{1}{\sqrt{2}} (\alpha_j^\dagger
\tilde{\alpha}_j)^{(1)}_M\right]
\; , \\
P_{2M} &\longleftrightarrow& 2\left[(d^\dagger s+s^\dagger
\tilde{d})^{(2)}_M
     - (\alpha_j^\dagger \tilde{\alpha}_j)^{(2)}_M\right]
\; , \\
P_{3M} &\longleftrightarrow& -2\sqrt{2}\left[(d^\dagger
\tilde{d})^{(3)}_M
     + \frac{1}{\sqrt{2}} (\alpha_j^\dagger
\tilde{\alpha}_j)^{(3)}_M\right]
\; .
\end{eqnarray}
\end{mathletters}%
In the boson-fermion multipole operators (\ref{u64}) we recognize
immediately
the generators of the Spin$_{\rm BF}$(6) group applied in the
phenomenological IBFM boson-fermion dynamical symmetry
U$_{\rm B}\otimes$U$_{\rm F}$(4) \cite{IK81,BBI81}.
Similarly, from Eqs.\ (\ref{so8sinim}) we identify single
nucleon
transfer operators in the form used in the above phenomenological
dynamical symmetry, provided that we limit ourselves to $n=0,1$
states only and neglect the $B^\dagger B$ type terms:
\begin{equation}\label{transf}
a^+_{jm} \longleftrightarrow \alpha^{\dagger}_{jm}
+\textstyle{\frac{1}{\sqrt{2}}}\left[s^\dagger\tilde{\alpha}_{jm}
+\sqrt{5}(d^\dagger\tilde{\alpha}_{j})^{(j)}_m\right] \; .
\end{equation}

Clearly, once we construct a fermion
dynamical symmetry Hamiltonian from the Casimir operators of the
subgroups
appearing in the chain
\begin{equation}\label{so6ch}
{\rm SO}(8)\supset {\rm Spin}(6) \supset {\rm Spin}(5) \supset
{\rm Spin}(3) \; ,
\end{equation}
its boson-fermion image would be just the
IBFM U$_{\rm B}\otimes$U$_{\rm F}$(4) dynamical symmetry
Hamiltonian.
Since the original SO(8) Hamiltonian does not distinguish
even and odd systems, the mapped Hamiltonian must also
be the same for both even and odd system. Therefore, as already
argued in the SU(2) case, the classification
scheme in the boson-fermion space should be extended by embedding
the boson-fermion dynamical symmetry into the U(6/4) dynamical
supersymmetry.  However, there is a difference between the mapped
boson-fermion system
and the phenomenological IBFM U(6/4) supersymmetry.  In the
SO(8) model considered here all particles occupy the same $j$=3/2
level while in the IBFM it
is assumed that the bosons occupy the whole valence shell with
only the fermion restricted to $j$=3/2.

As a more realistic situation, consider $k$=2, corresponding to
$j$=1/2, 3/2, 5/2, and 7/2.  In the nuclear shell model this
corresponds to the 3s$_{1/2}$, 2d$_{3/2}$, 2d$_{5/2}$, and 1g$_{7/2}$
orbitals located between the $N$=50 and 82 nuclear magic numbers.  In
the IBFM, a related supersymmetry with the same single-particle
content, is U(6/20), realized in the Au-Pt isotopes\cite{L84}.  The
group reduction chain is ${\rm U_{\rm B}(6)} \otimes {\rm U_F(20)} \supset
{\rm SO_B(6)} \otimes {\rm SU_F(4)} \supset {\rm Spin_{BF}(6)}
\supset {\rm Spin_{BF}(5)} \supset {\rm Spin_{BF}(3)}$.  A
Hamiltonian chosen as a linear combination of quadratic Casimir
operators appearing in this chain yields the analytical U(6/20) IBFM
energy formula \cite{L84}
\begin{eqnarray}
E &=& A \sigma(\sigma+4) +
    \tilde{A} [\sigma_1 (\sigma_1
+4)+\sigma_2(\sigma_2+2)+\sigma_3^2]
                                  \nonumber \\
    && + B[\tau_1(\tau_1+3)+\tau_2(\tau_2+1)]
    + C J(J+1) . \label{eqsy7}
\end{eqnarray}

In the simple extension of the SO(8) Ginocchio model to odd systems
the inactive angular momentum $k$ of the odd fermion gives rise to an
unrealistic degeneracy.  One can, however, lift this degeneracy by
adding to the SO(8) algebra multipole operators corresponding to an
interchange of the active and inactive angular momenta $k$ and $i$.
Suppose we add two such operators $\bar{P}_J$ for $J$=1 and 3,
\widetext
\begin{equation}\label{oddmult}
   \bar{P}_{JM} =
   -\sqrt{2\Omega} \sum_{j_1j_2}(-1)^{J+i+k+j_2}
   \frac{\hat{j_1}\hat{j_2}}{\hat{i}}
   \left\{\begin{array}{ccc} j_1 & j_2 & J \\
                             k   & k     & i
\end{array}\right\}
   \left(a^+_{j_1}\tilde{a}_{j_2}\right)^{(J)}_{M} , \label{oddp}
\end{equation}
\narrowtext
with  $k=2$ and $i=\textstyle{\frac{3}{2}}$.  These
operators form an SO(5) algebra and commute with the SO(8) generators
(\ref{so8gen}).  The resulting algebraic structure is therefore
SO(8)$\otimes$SO(5).  Hence, we can form a new dynamical symmetry
subgroup chain by combining the SO(5) generators with the Spin(5)
generators to obtain an additional group denoted by $\widetilde{\rm
Spin}$(5).  The fermion group reduction chain is then
\widetext
\begin{equation}\label{spin5ch}
{\rm SO(8)}\otimes {\rm SO(5)}\supset
{\rm Spin(6)}\otimes {\rm SO(5)}\supset {\rm Spin(5)}\otimes {\rm
SO(5)}
\supset \widetilde{\rm Spin}{\rm (5)}\supset \widetilde{\rm
Spin}{\rm (3)} \; ,
\end{equation}
\narrowtext
with ${\rm \widetilde{Spin}(5)}$
generated by
\begin{mathletters}
\label{til5gen}
\begin{eqnarray}
G_3 &=& P_3 - 2\sqrt{\textstyle{\frac{2}{5}}} \bar{P}_3 \; ,
\\
G_1 &=& P_1 + 2\sqrt{\textstyle{\frac{2}{5}}} \bar{P}_1 \; ,
\end{eqnarray}
\end{mathletters}%
and ${\rm
\widetilde{Spin}(3)}$ by $G_1$, with $P_J$  the original
SO(8)
multipole operators (\ref{so82}). We are now in a position to write
a Hamiltonian corresponding to the fermion dynamical symmetry
(\ref{spin5ch}) which is composed of quadratic Casimir operators
of the groups appearing in the chain:
\begin{equation}\label{hamsp5}
H = A C_{2 {\rm Spin(6)}} + B C_{2 {\rm Spin(5)}}
  + \tilde{B} C_{2 {\rm \widetilde{Spin}(5)}}
  + C C_{2 {\rm \widetilde{Spin}(3)}} \; ,
\end{equation}
with
\begin{mathletters}\label{casim}
\begin{eqnarray}
C_{2 {\rm Spin(6)}} &=&
{\textstyle{1\over 4}}(P_1\cdot P_1+P_2\cdot P_2+P_3\cdot P_3) \;
, \\
C_{2 {\rm Spin(5)}} &=&
{\textstyle{1\over 4}}(P_1\cdot P_1+P_3\cdot P_3) \; , \\
C_{2 {\rm \widetilde{Spin}(5)}} &=&
{\textstyle{1\over 4}}(G_1\cdot G_1+G_3\cdot G_3) \; , \\
C_{2 {\rm \widetilde{Spin}(3)}} &=&
\textstyle{\frac{5}{4}}G_1\cdot G_1 \; .
\end{eqnarray}
\end{mathletters}%
The corresponding  spectrum is obtained in a basis classified
by quantum numbers characterizing the representations of the
subgroups
in the dynamical symmetry chain, namely
\widetext
\begin{eqnarray}
E &=&   A [\sigma_1 (\sigma_1 +4)+\sigma_2(\sigma_2+2)+\sigma_3^2]
      + B[\tau_1(\tau_1+3)
                                        \nonumber \\
  & & + \tau_2(\tau_2+1)]
      + \tilde{B}[\tilde{\tau}_1(\tilde{\tau}_1+3)
    +\tilde{\tau}_2(\tilde{\tau}_2+1)]
    + C J(J+1) , \label{eqsy5}
\end{eqnarray}
\narrowtext
where $(\tau_1,\tau_2)$ are the ${\rm Spin(5)}$ irreps
and $(\tilde{\tau}_1,\tilde{\tau}_2)$ are the
irreps of ${\rm \widetilde{Spin}(5)}$.
For an even system comprised of $N$ fermion pairs, the quantum
numbers are:
$(\sigma_1,\sigma_2,\sigma_3)=(\sigma,0,0)$, with
$\sigma=N,N-2,N-4,\ldots 1,$ or $0$, $(\tau_1,\tau_2)=(\tau,0)$,
with
$\tau=0,1,\ldots \sigma$. One also has
$(\tilde{\tau}_1,\tilde{\tau}_2)=(\tau_1,\tau_2)$.
In the odd case, when the system is formed by $N$ fermion pairs
and a single odd fermion, the branching rules are
$(\sigma_1,\sigma_2,\sigma_3)=(\sigma\pm \textstyle{\frac{1}{2}},
\textstyle{\frac{1}{2}},\pm\textstyle{\frac{1}{2}})$, with
$\sigma=N,N-2,N-4,\ldots 1,$ or $0$,
$(\tau_1,\tau_2)=(\tau_1,\textstyle{\frac{1}{2}})$, with
$\sigma_1\ge\tau_1\ge\textstyle{\frac{1}{2}}$. Eventually, the
values of $(\tilde{\tau_1},\tilde{\tau_2})$,
$\tilde{\tau_1}\ge\tilde{\tau_2}$ are deduced from the reduction
pattern
\widetext
\begin{equation}\label{tildetau}
(\tau_1,\textstyle{\frac{1}{2}})
\otimes (1,0) = (\tau_1+1,\textstyle{\frac{1}{2}})
+(\tau_1,\textstyle{\frac{1}{2}})+
(\tau_1-1,\textstyle{\frac{1}{2}})+(\tau_1,\textstyle{\frac{3}{2}
}) \; .
\end{equation}
\narrowtext
In table\ \ref{tab1} we present the angular momentum $J$ content of
the lowest Spin(5) representations.

To arrive at an equivalent boson-fermion description we apply the
boson-fermion mapping (\ref{so8images}) and (\ref{so8sinim}),
whereas the new SO(5) algebra is simply mapped from the original
fermion space to the ideal fermion space by replacing the
operators $a_j$ by operators $\alpha_j$.
\begin{equation}\label{oddmultmap}
   \bar{P}_{JM} \longleftrightarrow
   -\sqrt{2\Omega} \sum_{j_1j_2}(-1)^{J+i+k+j_2}
   \frac{\hat{j_1}\hat{j_2}}{\hat{i}}
   \left\{\begin{array}{ccc} j_1 & j_2 & J \\
                             k   & k     & i
\end{array}\right\}
   \left(\alpha^{\dagger}_{j_1}\tilde{\alpha}_{j_2}\right)^{(J)}_{M}
\; .
\end{equation}
Obviously, the boson-fermion images of the generators of ${\rm
\widetilde{Spin}(5)}$ are obtained by combining the images
of Spin(5) and SO(5), as in expressions (\ref{til5gen}).
It can be verified that, up to a normalization factor, these
are just
the generators of the subgroup Spin$_{\rm BF}$(5) of the IBFM
U(6/20) (\ref{eqsy7}) given explicitly in Ref.
\cite{ND88}.  Consequently, at the boson-fermion level we have
now the following group chain
\widetext
\begin{eqnarray}\label{boschain}
&&{\rm U_B(6)} \otimes {\rm
U_F(20)} \supset {\rm U_B(6)} \otimes {\rm U_F(4)} \otimes {\rm
U_F(5)} \supset {\rm SO_B(6)} \otimes {\rm SU_F(4)} \otimes {\rm
SO_F(5)} \nonumber \\
&&\supset {\rm Spin_{BF}(6)} \otimes {\rm SO_F(5)}
\supset {\rm Spin_{BF}(5)} \otimes {\rm SO_F(5)} \supset {\rm
\widetilde{Spin}_{BF}(5)}
\supset {\rm \widetilde{Spin}_{BF}(3)} \; .
\end{eqnarray}
\narrowtext
The groups Spin$_{\rm BF}$($l$) correspond to Spin($l$) and are
generated
by the images (\ref{eqso87}), while the groups ${\rm
\widetilde{Spin}_{BF}({\it l})}$ correspond to ${\rm
\widetilde{Spin}({\it l})}$ and their generators are obtained as
outlined above.

It is interesting to note that the construction (\ref{oddmult}) 
and its implementation
in the group chain (\ref{spin5ch}) lead to 
a Hamiltonian (\ref{hamsp5}) with spectrum (\ref{eqsy5}) which is
the same as that used by Jolie {\it et al} \cite{Jol87}, but motivated from a
different point of view. The construction in Ref.\ \cite{Jol87} introduces an
{\it ad hoc} coupling between a pseudo-orbital angular momentum 2 and
pseudospin 3/2 to obtain the appropriate set of single-particle orbits for
their analysis of the A=130 mass region which is furthermore based on the
introduction of phenomenological s- and d-bosons to establish a U(6/20)
supersymmetry framework. Our construction therefore implies that the analysis
of Ref.\ \cite{Jol87} can be linked to and equivalently carried out in an 
SO(8)$\otimes$SO(5) framework, which becomes even more transparent after we 
introduce below the boson-fermion image of the Hamiltonian (\ref{hamsp5}).

To obtain the boson-fermion image of the Hamiltonian
(\ref{hamsp5}),
all the generators are simply replaced by their boson-fermion images.
The eigenenergy expression remains the same as in (\ref{eqsy5}).  Also
the representation reduction pattern remains the same as in the
original fermion system, provided the boson number satisfies $N_{\rm
B}=N$ and the number of ideal fermions is either 0, or 1.
Consequently, for the present model application, there is a one-to-one
correspondence between fermion states and states in the
boson-plus-one-fermion space, so that spurious states
do not appear.
Furthermore, following the same reasoning as in Sec.\
\ref{sec3}, since the fermion interaction strengths pertain to even
and odd systems alike, boson-fermion eigenstates of the mapped
Hamiltonian are contained in a single supersymmetric representation
of U(6/20).

As a simple application of the supersymmetric ${\rm
SO(8)}\otimes{\rm SO(5)}$ energy formula (\ref{eqsy5}),
we compare
the corresponding even and odd spectra
 in Fig.\ \ref{fig1}.
The Hamiltonian parameters $B=35$ keV, ${\tilde B}=12.63$ keV,
and $C=18$ keV, are chosen so that the even part coincides with
the
one of Ref.\ \protect{\cite{L84}}. In the odd spectrum we then
find more low-lying
$J$=$\left(\textstyle{\frac{3}{2}}\right)$$^+$
states than one gets from the
phenomenological IBFM expression (\ref{eqsy7}). The dotted lines
connect states belonging to the same ${\rm \widetilde{Spin}(5)}$
representations. We observe that these representations are
ordered
with increasing energy as
$(\tilde{\tau}_1,\tilde{\tau}_2)$=
$(\textstyle{\frac{1}{2}},\textstyle{\frac{1}{2}})$,
$(\textstyle{\frac{3}{2}},\textstyle{\frac{1}{2}})$,
$(\textstyle{\frac{1}{2}},\textstyle{\frac{1}{2}})$,
$(\textstyle{\frac{3}{2}},\textstyle{\frac{1}{2}})$,
$(\textstyle{\frac{3}{2}},\textstyle{\frac{3}{2}})$,
$(\textstyle{\frac{5}{2}},\textstyle{\frac{1}{2}})$.
On the other hand,
the phenomenological IBFM U(6/20) supersymmetry (\ref{eqsy7})
gives,
approximately, the ordering of Spin$_{\rm BF}$(5)
representations as follows:
$(\tau_1,\tau_2)$=
$(\textstyle{\frac{1}{2}},\textstyle{\frac{1}{2}})$,
$(\textstyle{\frac{3}{2}},\textstyle{\frac{1}{2}})$,
$(\textstyle{\frac{5}{2}},\textstyle{\frac{1}{2}})$,
$(\textstyle{\frac{3}{2}},\textstyle{\frac{1}{2}})$,
$(\textstyle{\frac{3}{2}},\textstyle{\frac{3}{2}})$,
$(\textstyle{\frac{1}{2}},\textstyle{\frac{1}{2}})$.
A different ordering of representations also changes the selection
rules of the allowed transitions.  For example, the observed
$\left(\textstyle{\frac{5}{2}}\right)^+_2
\rightarrow\left(\textstyle{\frac{3}{2}}\right)^+_1$
transition in $^{195}$Au \cite{F94} is forbidden in U(6/20)
(\ref{eqsy7}) as it violates the
$\Delta\tau_1$=0, $\pm 1$ selection
rule.  In the scheme presented here the
$\left(\textstyle{\frac{5}{2}}\right)^+_2$
state belongs to the
$(\textstyle{\frac{3}{2}},\textstyle{\frac{1}{2}})$ representation and
the above transition would be allowed.  However, it is at the moment
too speculative to claim that the dynamical supersymmetry discussed
here may be realized in the Pt-Au region 
and allow such selection rules to be tested.
It is interesting, though, to note that the discussed 
${\rm SO(8)}\otimes{\rm SO(5)}$ scheme, which we introduced in Ref.\
\cite{NGD94}, was also used recently in Ref.\ \cite{PPF96} and 
appears to have experimental evidence in the Xe-Ba region.
See e.g.\ Fig.\ ~3 of Ref.\
\cite{PPF96} with more experimental
low-lying low spin states in the odd systems than seems typical for
phenomenological supersymmetric spectra derived 
from (\ref{eqsy7}).

We have discussed the cases of one or zero unpaired fermions only.
However, it may be shown that a one-to-one correspondence between the
fermion states and the boson-fermion states holds when more unpaired
particles are considered as well, provided that the ideal fermions do
not form collective SO(8) pairs and, moreover, provided that the
condition $2N_{\rm B}+n\leq \Omega$ is satisfied.

\section{Conclusion}
\label{sec5}

In summary, we have applied
the recently constructed generalized Dyson boson-fermion
mapping of collective algebras to
the seniority SU(2) and SO(8)$\otimes$SO(5) models.
The mapping gives finite non-hermitian boson-fermion images of
collective pairs and single fermion operators expressed in terms of
ideal boson and fermion annihilation and creation operators.  In these
models, and for the specific interactions chosen, we have revealed a
dynamical supersymmetric structure analogous to what is found in the
interacting boson-fermion model, but with full recognition of the
Pauli principle.

The findings discussed in this paper are based on the construction of
a Hamiltonian from the generators of the fermion spectrum
generating
algebra which defines a class of dynamical symmetries\cite{SGA88}. The
Hamiltonian itself is not invariant with respect to the symmetry
group, but its eigenstates can be classified by the group and subgroup
representations.  At this level the supersymmetric structure is not
readily visible, and even and odd states simply belong to different
representations of the fermion group.  However, we showed that an
equivalent description which does reveal a dynamical supersymmetry can
be obtained after a boson-fermion mapping.

Whereas the Hamiltonian in the SU(2) model was constructed from the
pair operators which are mapped in a non-hermitian way (but
nevertheless yields a hermitian image for the pairing Hamiltonian), in
the SO(8) model application the Hamiltonian consisted of multipole
operators with manifestly hermitian boson-fermion images.
Apparently, a dynamical supersymmetric construction like the one
applied to the SO(8) model could also be extended to any fermion
algebraic model based on a pseudo-spin and pseudo-orbital recoupling,
as e.g.\ the SD Sp(6) \cite{Gin80}
and SDG SO(12) \cite{NGDD95} models. In such cases the
unification of the collective and non-collective algebras can always
be achieved on the Spin(3) level.  In the SO(8) example the
corresponding group was denoted by ${\rm \widetilde{Spin}(3)}$.  The
interesting question, however, is if one can find a unification of
some of the groups relevant to a given model at some higher level, as
in the case of ${\rm \widetilde{Spin}(5)}$ for the SO(8) example.
This would be analogous to a dynamical supersymmetry, or a
boson-fermion dynamical symmetry in the usual IBFM sense, where the
unification is usually considered to be formed at the highest level
possible in the group chain.

Obviously, a fermion algebraic model puts stronger
restrictions on the possible extensions to odd systems.  In any case,
should an algebraic model be applied to describe even-even nuclei,
like e.g.\ the fermion dynamical symmetry model (FDSM) \cite{fdsm},
its extension to the description of odd nuclei should also be
investigated.
As we have shown in this paper, the identification
of dynamical symmetries (the usual starting point of such analyses)
can sometimes be incorporated into a framework of dynamical
supersymmetries after the boson-fermion mapping, thus unveiling a
possibly richer dynamical symmetry structure than is immediately
visible on the original fermion level. It is of interest to enquire
which other ``hidden" non-trivial dynamical supersymmetries exist in
algebraic fermion models and to understand the relationship between
representations on the fermion level and those on the boson-fermion
level from a purely algebraic point of view, without necessarily
relying on the boson-fermion type mapping considered here.

Let us finally remark that the dynamical supersymmetry discussed in
the present context presupposes that all ideal space states enter the
analysis, as in the case of the SO(8) and SO(8)$\otimes$SO(5) models.
If some of these states do turn out to be spurious in other cases, as
they most likely will, one will have a situation which is the
equivalent of what is termed the ``dynamical Pauli effect" in the FDSM
\cite{fdsm,DNG94,dKG91}.  An incomplete supersymmetric spectrum will
result, but the classification and description of the remaining
physical states in terms of (ideal) bosons and fermions will still be
perfectly in order.

\acknowledgements{ This work is supported by the NSF grant No.
PHY93-21668 and grants from the Foundation for Research Development of
South Africa, the University of Stellenbosch, and in part by the
Polish State Committee for Scientific Research under Contract
No.~2~P03B~034~08
and by the Czech Republic grants GA ASCR A1048504 and GA
CR No. 202/93/2472.  HBG also acknowledges support in part by the
National Science Foundation under Grant No.\ PHY94-07194 and the
hospitality of the ITP at Santa Barbara during final completion of the
manuscript. }

\appendix
\section*{}
In order to prove that the boson-fermion mapping (\ref{e228})
preserves the commutation and anticommutation relations from
the original fermion space we proceed
in two steps.
First we use the boson-fermion mapping,
\begin{mathletters}
\label{e128}\begin{eqnarray}
   A^j               &\longleftrightarrow& R^j \equiv
                       {\cal{A}}^j  + B^i\big[{\cal{A}}_i,
                       {\cal{A}}^j\big]
                     - \textstyle{\frac{1}{2}}c^{jl}_{ik}B^i B^k
B_l
                                           , \label{e128c} \\
   \big[A_i,\!A^j\big] &\longleftrightarrow&
                            \big[{\cal{A}}_i,{\cal{A}}^j\big]
                    - c^{jl}_{ik}B^kB_l
                                              , \label{e128b} \\
   A_j               &\longleftrightarrow&
                            B_j               , \label{e128a} \\
   a^{\nu}           &\longleftrightarrow&
                            \alpha^{\nu}
                    + B^i\big[{\cal{A}}_i,\alpha^{\nu}\big]
                                              ,  \label{e128e} \\
   a_{\nu}           &\longleftrightarrow&
                            \alpha_{\nu}      ,  \label{e128d}
   \end{eqnarray}
\end{mathletters}%
 derived in Ref.~\cite{DSG93}    using
  supercoherent states.
This mapping preserves {\em exactly}
the commutation relations of the
collective algebra, as well as the commutation relations
between single-fermion and pair operators, and also the
anticommutation relations between single-fermion operators.  However,
a repeated application of the image of the pair
creation operator (\ref{e128c}) fails to produce
an ideal even space
spanned by bosons only.
(See also Ref.\ \cite{NGD95}.)

Second, to prove that the similarity
transformation, transforming the
mapping
(\ref{e128}) into the desired mapping
(\ref{e228}), is given by Eq.~(\ref{eq3}),
we proceed as follows.
We begin by noticing that $X^{-1}B_jX$=$B_j$,
because $X$ explicitly commutes with
the
boson annihilation operator $B_j$, accounting for (\ref{e228a}).
Next we show that $X^{-1}R^jX$=$R^j$$-$${\cal{A}}^j$ by
proving the
identity
\begin{equation}\label{e4}
\left[X,R^j-{\cal{A}}^j\right]={\cal{A}}^jX \;  .
\end{equation}
This is achieved by splitting $X$ into an infinite sum of terms
which individually increase the number of ideal fermions by 0, 2,
4,$\ldots$,
i.e., $X$=$\sum_{n=0}^{\infty}X_n$. As $R^j$$-$${\cal{A}}^j$
must conserve the number of ideal fermions,
from Eq.\ (\ref{e4})
we find the recurrence relation
$   \left[X_{n+1},R^j-{\cal{A}}^j\right]={\cal{A}}^jX_n $
which allows all $X_n$ to be determined, if $X_0$ is known.

We are free to choose $X_0$ as an arbitrary operator which
commutes
with $R^j$$-$${\cal{A}}^j$, and Eq.\ (\ref{eq3}) reflects the
simplest choice $X_0$=1.  Then the term $X_1$
is a solution of the recurrence relation for $n$=0.  Indeed we
have
\widetext
   \begin{eqnarray}\label{e6}
       \bigg[\frac{1}{C_{\rm F}-\widehat{C}_{\rm F}} {\cal A}^i B_i\:\:
         {\textstyle \raisebox{-2ex}{$\widehat{}$}}\:\:,
         R^j-{\cal{A}}^j\bigg]
  & = &\frac{1}{C_{\rm F}-\widehat{C}_{\rm F}} \left[{\cal A}^i B_i,
         R^j-{\cal{A}}^j\right]\:
         {\textstyle \raisebox{-2ex}{$\widehat{}$}} \nonumber \\
  & = &\frac{1}{C_{\rm F}-\widehat{C}_{\rm F}}
(C_{\rm F}{\cal{A}}^j-{\cal{A}}^jC_{\rm F})\:
         {\textstyle \raisebox{-2ex}{$\widehat{}$}} =
{\cal{A}}^j,
   \end{eqnarray}
where the first equality results from the fact that $C_{\rm F}$
commutes with $R^j$$-$${\cal{A}}^j$ and the second one from the
explicit calculation of the commutator. Similarly, we prove by
induction that the $X_{n+1}$ term of Eq.\ (\ref{eq3}) is a
solution of the recurrence relation provided the same is true for
$X_n$:
   \begin{eqnarray}\label{e7}
       [X_{n+1} , R^j-{\cal{A}}^j]
    &=& \frac{1}{C_{\rm F}-\widehat{C}_{\rm F}} \left[{\cal{A}}^i B_iX_n,
         R^j-{\cal{A}}^j\right]\:\:
         {\textstyle \raisebox{-2ex}{$\widehat{}$}}\:\: \nonumber
\\
    &=& \frac{1}{C_{\rm F}-\widehat{C}_{\rm F}} \left({\cal{A}}^i B_i
        {\cal{A}}^jX_{n-1}
         + (C_{\rm F}{\cal{A}}^j-{\cal{A}}^jC_{\rm F})X_n\right)\:\:
         {\textstyle \raisebox{-2ex}{$\widehat{}$}} \nonumber \\
    &=& \:{\textstyle \raisebox{-1ex}{$\breve{}$}}\:\:{\cal{A}}^j
       \frac{1}{\breve{C_{\rm F}}-\widehat{C}_{\rm F}}\bigl(1
      +\frac{\breve{C_{\rm F}}-C_{\rm F}}{C_{\rm F}-\widehat{C}_{\rm F}}\bigr)
       {\cal{A}}^i B_iX_{n-1}\:\:
         {\textstyle \raisebox{-2ex}{$\widehat{}$}}\:\: =
       {\cal{A}}^jX_n
   \end{eqnarray}
\narrowtext
where the operator $\breve{C_{\rm F}}$ is in turn evaluated at
``$\:$\raisebox{-1ex}{$\breve{}$}$\:$''.

Finally, to see that Eq. (\ref{e228b}) is invariant under the
transformation
(\ref{eq3}) we use the facts that
$\left[C_{\rm F},[{\cal{A}}_i,{\cal{A}}^j]\right]$=0
and $\left[{\cal{A}}^m B_m,\big[{\cal{A}}_i,{\cal{A}}^j\big]
                    - c^{jl}_{ik}B^kB_l\right]=0$ which follow
from the algebraic closure and the Jacobi identities.

\begin{figure}
\caption{The lower part of an SO(8)$\otimes$SO(5) spectrum.
In the odd part (right), the (1/2,1/2) and (3/2,1/2) irreps of
Spin(5) are shown only. The dotted lines connect states belonging
to the same $\tilde{\rm Spin}$(5) irrep.
}
\label{fig1}
\end{figure}

\begin{table}
\begin{tabular}{c|l}
$(\tau_1,\tau_2)$&$J$ \\
\hline
$(0,0)$&0 \\
$(1,0)$&2  \\
$(2,0)$&4,2 \\
$(3,0)$&6,4,3,0 \\
$(4,0)$&8,6,5,4,2 \\
\hline
$(\textstyle{\frac{1}{2}},\textstyle{\frac{1}{2}})$&$\textstyle{
\frac{3}{2}}$ \\
$(\textstyle{\frac{3}{2}},\textstyle{\frac{1}{2}})$&
$\textstyle{\frac{7}{2}},\textstyle{\frac{5}{2}},\textstyle{\frac
{1}{2}}$ \\
$(\textstyle{\frac{5}{2}},\textstyle{\frac{1}{2}})$&
$\textstyle{\frac{11}{2}},\textstyle{\frac{9}{2}},\textstyle
  {\frac{7}{2}},
\textstyle{\frac{5}{2}},\textstyle{\frac{3}{2}}$ \\
$(\textstyle{\frac{3}{2}},\textstyle{\frac{3}{2}})$&
$\textstyle{\frac{9}{2}},\textstyle{\frac{5}{2}},\textstyle{\frac
{3}{2}}$
\end{tabular}
\caption{Angular momentum content of the lowest Spin(5)
representations
is presented. In the upper (lower) part the representations
relevant to the
even (odd) system are shown, respectively.
}
\label{tab1}
\end{table}

\end{document}